\documentclass[12pt]{article}
\usepackage{amsmath,cite,epsfig,a4,times,euscript,listings}
\usepackage[text-hat-accent]{euler}
\usepackage{graphics}
\usepackage[usenames]{color}
\usepackage[T1]{fontenc}
\usepackage{bold-extra}
\usepackage[nottoc,numbib]{tocbibind}
\renewcommand{\baselinestretch}{1.1}
\newcommand{\myTitle}[1]{\begin{center}{\bf\Huge #1}\\[5ex]\end{center}}
\newcommand{\myAuthor}[1]{\begin{center}{\Large #1}\\[2ex]\end{center}}
\newcommand{\myAffiliation}[1]{\\[1ex]{\it\large #1}}
\newcommand{\myEmail}[1]{}
\newcommand{\myDate}{\vspace{5ex}}
\newcommand{\myAbstract}[1]{\begin{center}\renewcommand{\baselinestretch}{1}{\bf Abstract}\\[2ex]\parbox{0.8\linewidth}{\small\hspace{15pt} #1}\end{center}\vspace{\baselineskip}}
\newcommand{\myReport}[1]{\hspace{\fill} #1}
\newcommand{\myPreprint}[1]{}
\newcommand{\myKeywords}[1]{}

\newcommand{\myScript}[1]{\EuScript{#1}}

\newcommand{\Appendix}[1]{Appendix~\ref{#1}}   
\newcommand{\Section}[1]{Section~\ref{#1}}

\newcommand{\Equation}[1]{Eq.~(\ref{#1})}

\newcommand{\ie}{{\it i.e.}}

\newcommand{\eg}{{\it e.g.}}
\newcommand{\KaTie}{{\sc Ka\nolinebreak\hspace{-0.3ex}Tie}}
\newcommand{\avhlib}{{\sc avhlib}}
\newcommand{\mytt}[1]{{\small\tt#1}}
\newcommand{\myvm}[1]{\\[1ex]\noindent{\small\tt#1}\\[1ex]\noindent}
\newcommand{\flux}{\mathrm{flux}}%
\newcommand{\Wmax}{W_{\mathrm{max}}}%
\newcommand{\Wthrs}{W_{\mathrm{thrs}}}%
\newcommand{\NMPI}{N_\mathrm{MPI}}%
\newcommand{\ubar}{\bar{u}}%
\newcommand{\dbar}{\bar{d}}%
\newcommand{\sbar}{\bar{s}}%
\newcommand{\cbar}{\bar{c}}%
\newcommand{\bbar}{\bar{b}}%
\newcommand{\qBAR}{q{\scriptstyle\sim}}%
\newcommand{\uBAR}{u{\scriptstyle\sim}}%
\newcommand{\dBAR}{d{\scriptstyle\sim}}%
\newcommand{\lhs}[2]{{\tt#1\hspace{1ex}#2}}
\newcommand{\Esoft}{E_{\mathrm{soft}}}%
\newcommand{\Ecm}{E_{\mathrm{cm}}}%

\newcommand{\tweakcodepar}[3]%
  {\vspace{#1ex}\newline\noindent\hspace*{4.0ex}{\small\tt #3}\vspace{#2ex}\newline\noindent}

\begin{document}

\myReport{IFJPAN-IV-2016-26}
\myPreprint{}\\[2ex]

\myTitle{%
\textsc{Ka\hspace{-0.3ex}Tie}:
for parton-level event generation\\[0.5ex]
with ${\bf k_T}$-dependent initial states
}

\myAuthor{%
A.~van~Hameren%
\myAffiliation{%
Institute of Nuclear Physics Polish Academy of Sciences\\
PL-31-342 Krak\'ow, Poland%
\myEmail{hameren@ifj.edu.pl}
}
}

\myDate

\myAbstract{%
\KaTie\ is a parton-level event generator for hadron scattering processes that can deal with partonic initial-state momenta with an explicit transverse momentum dependence causing them to be space-like.
Provided with the necessary transverse momentum dependent parton density functions, it calculates the tree-level off-shell matrix elements and performs the phase space importance sampling to produce weighted events, for example in the Les Houches Event File format.
It can deal with arbitrary processes within the Standard Model, for up to at least four final-state particles.
Furthermore, it can produce events for single-parton scattering as well as for multi-parton scattering.
}
\newpage

\myKeywords{QCD, Monte Carlo, off-shell, tree-level, Standard Model}

%\begin{document}
%
\tableofcontents
\section{Introduction}
Event generators are indispensable for the scientific activity in high-energy physics.
Scattering experiments performed at for example the Large Hadron Collider involve physical processes evolving at a large range of energy scales.
Typical energy scales reach several TeV in the so-called hard scattering process, in which heavy elementary particles like the Higgs boson are created, and evolve down to below the level of a GeV when the eventual scattering remnants reach the detector.
Multi-purpose event generators%
~\cite{%
Sjostrand:2014zea%
,Bahr:2008pv%
%,Bellm:2015jjp%
,Gleisberg:2008ta%
}%
\ are designed to simulate scattering process over the whole range of scales.
Their development greatly benefited from dedicated studies of the parton-level hard scattering process, which is computationally accessible with perturbative quantum chromodynamics (QCD).
The goal of computing cross sections for arbitrary tree-level processes with many final-state particles (four or more) led to creation of several programs around the year 2000~\cite{Kanaki:2000ey,Krauss:2001iv,Mangano:2002ea,Moretti:2001zz,Kilian:2007gr,Maltoni:2002qb}.
The descriptive power of both the multi-purpose event generators and the parton-level event generators was greatly enhanced by combining their efforts using so-called merging procedures~\cite{Buckley:2011ms,Sjostrand:2016bif}.
The challenge of the parton-level generators was twofold: phase space points must be generated efficiently, and matrix elements must be evaluated quickly.
The program presented in this paper is of this type, but with an extended range of applicability, as will be explained below.
It must be mentioned that developments have gone beyond tree-level.
Several parton-level event generators operate at next-to-leading (NLO) precision \cite{Frixione:2008ym,Arnold:2008rz,Campbell:2010ff,Alioli:2010xd,Bevilacqua:2011xh,Alwall:2014hca,Weiss:2015npa}.
Obtaining the necessary one-loop amplitudes was a bottleneck, and several programs dedicated to solve this were developed \cite{Berger:2008sj,Cascioli:2011va,Cullen:2011ac,Badger:2012pg,Actis:2016mpe}.
The multi-purpose generators have been highly advanced, steering full calculations using internal libraries for hard scattering matrix elements or using external programs in various combinations, leading to precise predictions, \eg~\cite{%
Badger:2013ava%NJet
,vanDeurzen:2013xla%GoSam
,Bern:2014vza%BlackHat
,Kallweit:2015dum%OpenLoops
,Rauch:2016upa,Bellm:2016cks%
}.

All these programs function within a factorization prescription that separates the low-scale physics of the colliding protons from the high-scale physics of the hard process.
More specifically, the cross section is a convolution of the parton distribution functions (PDFs) describing the protons, and the matrix element, describing the hard process.
The PDFs only depend on the factorization scale and the fraction of the light-like proton momenta that enters the hard process, as prescribed by {\em collinear factorization\/}~\cite{%
Libby:1978qf%
,Ellis:1978ty%
,Collins:1989gx%
}.

The initial-state partons in the hard process being parallel to the incoming hadrons is an approximation.
For example, it implies that the final-state momenta can be separated into two groups that are back-to-back, which in a realistic collision is never the case.
This approximation is traditionally cured by including higher fixed-order corrections, and/or by augmentation with a parton shower.
It was argued in~\cite{Collins:2005uv} that it would be desirable to remove this approximation already at lowest order in perturbation theory, and allow for non-vanishing transverse momentum components of the initial-state momenta in the hard process.
The higher-order corrections may become much smaller, and it would open the possibility to include resummation corrections via transverse momentum dependent (TMD), or {\em un-integrated}, PDFs.
An example of such an approach is high-energy factorization ($k_T$-factorization)~\cite{Catani:1990eg,Collins:1991ty,Levin:1991ry}.

On the side of the TMDs necessary for such an approach, there has been substantial activity~\cite{Angeles-Martinez:2015sea}.
Several evolution equations for TMDs have been developed, and  the CCFM evolution~\cite{Ciafaloni:1987ur,Catani:1989yc,Catani:1989sg,Marchesini:1994wr} in particular is employed in the event generator {\sc CASCADE}~\cite{Jung:2010si}.
Several TMDs are provided by the library {\sc TMDlib}~\cite{Hautmann:2014kza}.
Besides the TMDs, such calculation requires matrix elements with off-shell initial-state partons.
These cannot be obtained from collinear matrix elements by just changing the kinematics, because that would in general break gauge invariance.
The problem of defining and calculating them, at least at tree-level, has been solved~\cite{%
Lipatov:1995pn%
,Lipatov:2000se%
,Antonov:2004hh%
,vanHameren:2012if%
,vanHameren:2013csa%
,vanHameren:2013gba%
,Kotko:2014aba%
},
and an automated implementation for arbitrary processes within the Standard Model with off-shell initial-state partons and with several final-state particles can be found in \avhlib~\cite{Bury:2015dla}.
Furthermore, the program {\sc OGIME}~\cite{Kotko:Ogime} can generate analytic expressions for multi-gluon amplitudes with several of them off-shell.

Besides implementations of functions to evaluate the TMDs and the matrix elements, the calculation of a cross section or other observables requires Monte Carlo tools to perform the necessary phase space integrals.
The program {\sc LxJet}~\cite{Kotko:LxJet} can be used for a selection of processes.
For arbitrary processes, the tools are publicly available in \avhlib~\cite{vanHameren:2007pt,vanHameren:2010gg}, which has already been used for calculations of various processes at the LHC, like for example $pp \to 4j$, $pp \to Z+j$ and $pp \to c\bar{c}\,c\bar{c}$ productions~\cite{%
Kutak:2016ukc%
,Kutak:2016mik%
,vanHameren:2015uia%
,vanHameren:2015wva%
}.

While \avhlib\ provides all the numerical tools to generate phase space points, interpolate PDF grids, evaluate matrix elements etc., it does not provide a practical environment to perform phenomenological studies.
\KaTie\ fills this hiatus.
In particular, it produces event files in the Les Houches Event File (LHEF) format~\cite{Alwall:2006yp}, or a custom format for which it provides the tools to produce distributions of arbitrary observables.
The definition of the process one would like to study happens via a single input file, containing all the information about which subprocesses are included, which PDFs are used, which cuts are applied, which values of model parameters are used, etc..
\KaTie\ does not provide any PDFs, but uses LHAPDF~\cite{Buckley:2014ana} for collinear PDFs, and automatically interpolates any TMD PDF provided in the form of a (hyper-)rectangular grid (further described in \Section{subsubPDFs}).
Alternatively, {\sc TMDlib} can be used to provide TMD PDFs.
Finally, \KaTie\ offers a convenient environment to perform calculations for multi-parton scattering (MPI), in which more than one hard process occurs in each event simultaneously.
%
%There is no correlation between the initial-state partons coming from the same proton in the events that \KaTie\ produces, but this can conveniently be included by means re-weighting.
%

The outline of the paper is as follows. 
\Section{formalism} introduces the formalism along with some notation, and the usage of the program is explained in \Section{Usage}.
\Section{Sec:summary} contains the summary, and the appendices containing some details close the paper.
\section{Formalism\label{formalism}}
We consider collision processes of specific hadrons resulting in a specific final state and refer to those with the symbol $Y$.
The process $Y=p\,p\to\mu^+\mu^-\,j\,j$ (proton-proton to $\mu^+\mu^-$ plus two jets) is a typical example.
Several partonic processes contribute in this example, \eg\ $Y\ni y=\bar{u}\,g\to\mu^+\mu^-\,\bar{u}\,g$.
The separate particles in the partonic process are refered to by $y_i$.
In the example given before, we have $y_2=g$ and $y_4=\mu^-$.
The number of final-state particles in $Y$ is $n=4$ in the given example.
In the following, we assume that all processes in $Y$ share the same kinematical situation, and that the mass of final-state particle $i$ is the same for every $y\in Y$.

A generally factorized formula for the differential cross section of a hadron collision process with a single parton-level scattering is given by
%
%%%%%%%%%%%%%%%%%%%%%%%%%%%%%%%%%%%%%%%%
\begin{equation}
  d\sigma_Y(p_1,p_2;k_3,\ldots,k_{2+n}) =
  \sum_{y\in Y}\int d^4k_1\,\myScript{P}_{y_1}(k_1) \int d^4k_2\,\myScript{P}_{y_2}(k_2)\,
  d\hat{\sigma}_y(k_1,k_2;k_3,\ldots,k_{2+n})
~.
\end{equation}
%%%%%%%%%%%%%%%%%%%%%%%%%%%%%%%%%%%%%%%%
%
It is differential in the final-state momenta $k_3,\ldots,k_{2+n}$.
The symbols $p_1,p_2$ refer to the momenta of the hadrons, while $k_1,k_2$ are the momenta of the initial-state partons.
In collinear factorization, the distributions denoted with $\myScript{P}$ are given by
%
%%%%%%%%%%%%%%%%%%%%%%%%%%%%%%%%%%%%%%%%
\begin{equation}
\myScript{P}_{y_i}(k_i) = \int\frac{dx_i}{x_i}\,f_{y_i}(x_i,\mu)\,\delta^4(k_i-x_ip_i)
~,
\end{equation}
%%%%%%%%%%%%%%%%%%%%%%%%%%%%%%%%%%%%%%%%
%
where $f_{y_i}$ is the collinear PDF for a parton of type $y_i$ coming from hadron $i$, and $\mu$ is the, possibly dynamical, factorization scale.
For $k_T$-factorization the distributions typically look like
%
%%%%%%%%%%%%%%%%%%%%%%%%%%%%%%%%%%%%%%%%
\begin{equation}
\myScript{P}_{y_i}(k_i) = \int\frac{d^2{\bf k}_{iT}}{\pi}\int\frac{dx_i}{x_i}\,\myScript{F}_{y_i}(x_i,|{\bf k}_{iT}|,\mu)\,\delta^4(k_i-x_ip_i-k_{iT})
~,
\end{equation}
%%%%%%%%%%%%%%%%%%%%%%%%%%%%%%%%%%%%%%%%
%
where now $\myScript{F}_{y_i}$ is the TMD or unintegrated PDF, and $k_{iT}$ is the embedding of ${\bf k}_{iT}$ in four-dimensional Minkowski space.
The partonic differential cross section can be dissected futher as follows:
%
%%%%%%%%%%%%%%%%%%%%%%%%%%%%%%%%%%%%%%%%
\begin{align}
d\hat{\sigma}_y(k_1,k_2;k_3,\ldots,k_{2+n})
&=
d\Phi_Y(k_1,k_2;k_3,\ldots,k_{2+n})\,\Theta_Y(k_3,\ldots,k_{2+n})
\\\notag
&\times\flux(k_1,k_2)\times
\myScript{S}_y\,|\myScript{M}_y(k_1,\ldots,k_{2+n})|^2
~,
\end{align}
%%%%%%%%%%%%%%%%%%%%%%%%%%%%%%%%%%%%%%%%
%
where $d\Phi_Y$ is the $(3n-4)$-dimensional differential phase space element for a final state with $n$ particles and masses dictated by $Y$:
%
%%%%%%%%%%%%%%%%%%%%%%%%%%%%%%%%%%%%%%%%
\begin{equation}
d\Phi_Y(k_1,k_2;k_3,\ldots,k_{2+n})
=
(2\pi)^{4-3n}\Bigg[\prod_{i=3}^{2+n}d^4k_i\,\delta_+(k_i^2-m_i^2)\Bigg]
\delta^4\big(k_1+k_2-k_3-\cdots-k_{2+n}\big)
~.
\end{equation}
%%%%%%%%%%%%%%%%%%%%%%%%%%%%%%%%%%%%%%%%
%
The (squared) matrix element $|\myScript{M}_y|^2$ includes the summation over spins and colors of all external particles.
It is turned into the average over color and spin degrees of freedom for the initial-state particles through the factor $\myScript{S}_y$, which also includes the symmetry factor for the final state.
The matrix elements used in \KaTie\ are tree-level matrix elements, where $k_1$ and $k_2$ can have non-vanishing transverse components, and contain singularities which need to be avoided.
This is established by phase space cuts $\Theta_Y$, typically consisting of minimum transverse momenta, maximum absolute rapidities, minimum distance between momenta in the two-dimenional space of rapidity and azimutal angle ($\Delta R$), etc. (examples of implemented cuts and scales can be found in \Section{kinematicsAndCuts}).
The flux factor, finally, includes the demand that the energy of the partonic process is positive:
%
%%%%%%%%%%%%%%%%%%%%%%%%%%%%%%%%%%%%%%%%
\begin{equation}
\flux(k_1,k_2)
=
\frac{\theta\big((k_1+k_2)^2\big)}{4\sqrt{({k_1}\!\cdot\!{k_2})^2-k_1^2k_2^2}}
~.
\label{fluxfactor}
\end{equation}
%%%%%%%%%%%%%%%%%%%%%%%%%%%%%%%%%%%%%%%%
%
The particular denominator is not prescribed by a specific factorization theorem, and is just an analytic continuation of the textbook formula for massive initial-state particles.
It is discussed further at the end of \Section{kinematicsAndCuts}.

\subsection{Event generation}
\KaTie\ creates event files consisting of a list of weighted phase space points such that
%
%%%%%%%%%%%%%%%%%%%%%%%%%%%%%%%%%%%%%%%%%
\begin{equation}
\frac{1}{N}\sum_{I=1}^NW_I\,\myScript{B}\Big(k_3^{(I)},\ldots,k_{2+n}^{(I)}\Big)
\approx
\int d\sigma_Y(p_1,p_2;k_3,\ldots,k_{2+n})\,\myScript{B}(k_3,\ldots,k_{2+n})
~.
\end{equation}
%%%%%%%%%%%%%%%%%%%%%%%%%%%%%%%%%%%%%%%%
%
The sum is over the events, and the approximation formally becomes an equality when the number of events $N$ goes to infinity.
$W_I$ is the weight of phase space point $I$, and $k_{j}^{(I)}$ is the $j$-th momentum in event $I$. 
$\myScript{B}$ is a test function, for example the bin $[z,z+\delta z]$ for a distribution of an observable $\myScript{O}$:
%
%%%%%%%%%%%%%%%%%%%%%%%%%%%%%%%%%%%%%%%%
\begin{equation}
\myScript{B}(k_3,\ldots,k_{2+n}) =
\theta( \myScript{O}\big(k_3,\ldots,k_{2+n}) - z\big)
\,\theta( z+\delta z - \myScript{O}\big(k_3,\ldots,k_{2+n}) \big)
~.
\end{equation}
%%%%%%%%%%%%%%%%%%%%%%%%%%%%%%%%%%%%%%%%
%
Of course, \KaTie\ also provides $k_1,k_2$ in the event file, and ``observables'' depending on these momenta can be studied too.

\subsubsection{Importance sampling}
\KaTie\ uses importance sampling to reduce the fluctuation of the weights.
Because of the multitude of possible PDFs and matrix elements causing the fluctuations, the importance sampling is partly achieved in an adaptive manner, and \KaTie\ employs an optimization phase before it starts generating the actual events.
For each partonic subprocess, a probability density $P_{y}$ is adaptively created with the aim that
%
%%%%%%%%%%%%%%%%%%%%%%%%%%%%%%%%%%%%%%%%
\begin{align}
\myScript{W}_y(k_1,\ldots,k_{2+n})
=
\frac{\myScript{P}_{y_1}(k_1)\,\myScript{P}_{y_2}(k_2)\,\flux(k_1,k_2)\,\myScript{S}_{y}\,
|\myScript{M}_{y}(k_1,\ldots,k_{2+n})|^2}
{P_{y}(k_1,\ldots,k_{2+n})}
\end{align}
%%%%%%%%%%%%%%%%%%%%%%%%%%%%%%%%%%%%%%%%
%
is as constant as possible.
The differential cross section can be written as
%
%%%%%%%%%%%%%%%%%%%%%%%%%%%%%%%%%%%%%%%%
\begin{equation}
d\sigma_Y
=
\sum_{y\in Y}dF_{y}(k_1,\ldots,k_{2+n})\,\Theta_Y(k_3\ldots,k_{2+n})
\,\myScript{W}_{y}(k_1,\ldots,k_{2+n})
~,
\label{s2eq1}
\end{equation}
%%%%%%%%%%%%%%%%%%%%%%%%%%%%%%%%%%%%%%%%
%
where we say that phase space points are generated following the distributions $F_{y}$ given by
%
%%%%%%%%%%%%%%%%%%%%%%%%%%%%%%%%%%%%%%%%
\begin{equation}
dF_{y}(k_1,\ldots,k_{2+n}) = d^4k_1\,d^4k_2\,d\Phi_Y(k_1,k_2;k_3,\ldots,k_{2+n})\,P_{y}(k_1,\ldots,k_{2+n})
~.
\end{equation}
%%%%%%%%%%%%%%%%%%%%%%%%%%%%%%%%%%%%%%%%
%
Integration over $k_1,k_2$ is understood in \Equation{s2eq1}.
The creation of the distributions $F_{y}$ leads to crude estimates of the partonic cross sections
%
%%%%%%%%%%%%%%%%%%%%%%%%%%%%%%%%%%%%%%%%
\begin{equation}
\sigma_{y} = \int dF_{y}(k_1,\ldots,k_{2+n})\,\Theta_Y(k_3\ldots,k_{2+n})
\,\myScript{W}_{y}(k_1,\ldots,k_{2+n})
~.
\end{equation}
%%%%%%%%%%%%%%%%%%%%%%%%%%%%%%%%%%%%%%%%
%
During the event generation, first a subprocess is chosen with relative probability $\sigma_{y}$, and then a phase space point is generated following the distribution $F_{y}$.
If the event does not satisfy the phase space cuts $\Theta_Y$, it is rejected, and else it is accepted with weight
%
%%%%%%%%%%%%%%%%%%%%%%%%%%%%%%%%%%%%%%%%
\begin{equation}
W = \myScript{W}_{y}(k_1,\ldots,k_{2+n})\,\frac{\sum_{y'\in Y}\sigma_{y'}}{\sigma_{y}}
~.
\end{equation}
%%%%%%%%%%%%%%%%%%%%%%%%%%%%%%%%%%%%%%%%
%
So each event automatically has a subprocess assigned to it.

\subsubsection{Un-weighting}
Despite the importance sampling, the event weights may still fluctuate wildly, and supplementary techniques need to be applied to reduce this behavior.
In the crude un-weighting method, the maximum weight $\Wmax$ is determined, and event $I$ is accepted with probability $W_I/\Wmax$.
The main advantage is that all events get the same weight $\Wmax$.
The disadvantage is that in practice $\Wmax$ can only be determined within the total sample of events, and generating extra events may lead to a new increased $\Wmax$, effectively reducing the total number of events.

\KaTie\ uses partial un-weighting to reduce the weight fluctuation while avoiding the problem mentioned above.
Using the first $N_0$ events, a weight $\Wthrs$ is determined, and from then on, events with weight $W_I\geq\Wthrs$ are accepted and keep their weight, while events with weight $W_I<\Wthrs$ are accepted with probability $W/\Wthrs$.
If accepted, the latter get weight $\Wthrs$.
The threshold $\Wthrs$ is estimated with the aim to reach (the order of) a desired number of accepted events after reaching a desired statistical precision for the estimate of the total cross section.
This way, a compromise can be chosen between accepting all events with wildly fluctuating weights, and accepting (possibly only very) few events with constant weight.
More details can be found in the next section.
It is important to mention that event files created with different values of $\Wthrs$ can be safely mixed without creating a bias, as long as they were created with {\em exactly the same set of distributions $F_{y}$\/}.

\subsection{TMD interpolation}
\KaTie\ does not include any $k_T$-dependent PDFs.
These can be provided by {\sc TMDlib}~\cite{Hautmann:2014kza}.
Alternatively, they can be provided in the form of data files consisting of (hyper) rectangular grids representing the PDFs.
The columns in the files must contain
\begin{equation}
\ln(x) \quad  \ln\big(|\mathbf{k}_T|^2\big) \quad x\myScript{F}(x,|\mathbf{k}_T|)
\end{equation}
or
\begin{equation}
\ln(x) \quad  \ln\big(|\mathbf{k}_T|^2\big) \quad \ln\big(\mu^2\big) \quad x\myScript{F}(x,|\mathbf{k}_T|,\mu)
\end{equation}
if the PDF also depends on the factorization scale.
\KaTie\ employs routines from \avhlib, which use straightforward multi-linear interpolation, to interpolate these grids itself.
As mentioned, the grids must be rectangular, but they do not need to be regular.

\subsection{Matrix elements}
The necessary matrix elements are provided by {\sc avhlib}.
They are evaluated via numerical Dyson-Schwinger recursion, which is implemented in a way very similar to {\sc Helac}~\cite{Kanaki:2000ey,Cafarella:2007pc}.
Before the start of a Monte Carlo run, a list is prepared for each process, encoding the vertex operations and propagator operations that have to be executed numerically for each phase space point in order to arrive at the value of the amplitude.
So the vertices and propagators etc.\ exist as compiled routines in the executable, whereas the amplitudes exist as such lists.
Color is treated in terms of color-ordered decomposition, more specifically in the color-connection decomposition as in {\sc Helac}.
This has the advantage that it naturally generalizes to arbitrary numbers of quark-antiquark pairs, but has the disadvantage that it does not automatically lead to the most efficient calculation of multi-gluon amplitudes.
Because of this, the maximum number of color pairs in a process for \KaTie\ to deal with comfortably is $6$ for now.

\subsection{MPI}
It is possible to generate event files with \KaTie\ in which each event is the result of two or more independent scatterings.
These can be useful if one wants to study multiple parton interactions (MPI)~\cite{Astalos:2015ivw}.
In this case we consider the process $Y$ to be separated into $\NMPI$ processes $Y^{(h)}$.
For example $Y=p\,p\to\mu^+\mu^-\,j\,j$ could be separated into $Y^{(1)}=p\,p\to\mu^+\mu^-$ and $Y^{(2)}=p\,p\to j\,j$.
The hard processes are imagined to originate from one and the same pair of hadrons of course, but for our notation the above separation is most convenient.
It has to be stated that \KaTie\ does not provide a list of possible separations, and that a list of desired separations has to be provided by the user.

\KaTie\ treats MPI by simply factorizing all distributions, the matrix element, and the phase space.
The only function that is not necessarily factorized is $\Theta_Y$ representing the phase space cuts.
Furthermore, there are extra restrictions requiring the sum of hadron momentum fractions coming from the same hadron to be smaller than one.
The differential cross section becomes
%
%%%%%%%%%%%%%%%%%%%%%%%%%%%%%%%%%%%%%%%%
\begin{align}
d\sigma_Y
&=
\frac{\myScript{S}_{\mathrm{MPI}}}{\sigma_{\mathrm{eff}}}
\prod_{h=1}^{\NMPI}
\Bigg[\sum_{y\in Y^{(h)}}
dF_{y}\big(k_1^{(h)},\ldots,k_{2+n_h}^{(h)}\big)
\,\myScript{W}_{y}\big(k_1^{(h)},\ldots,k_{2+n_h}^{(h)}\big)
\Bigg]
\\\notag&\times
  \Theta_Y(k_3\ldots,k_{2+n})
\,\theta\Bigg(1-\sum_{h=1}^{\NMPI}x_1^{(h)}\Bigg)
\,\theta\Bigg(1-\sum_{h=1}^{\NMPI}x_2^{(h)}\Bigg)
~,
\end{align}
%%%%%%%%%%%%%%%%%%%%%%%%%%%%%%%%%%%%%%%%
%
where $\sigma_{\mathrm{eff}}$ is some phenomenologically determined normalization with units of a surface to the power $\NMPI-1$, and $\myScript{S}_{\mathrm{MPI}}$ is the symmetry factor associated with the MPI.
For each set of, say $l$, identical $Y^{(h)}$, it contains a factor $1/l!$.

In the event generation, a subprocess $y^{(h)}$ is chosen for each $Y^{(h)}$ with relative probability
%
%%%%%%%%%%%%%%%%%%%%%%%%%%%%%%%%%%%%%%%%
\begin{equation}
\sigma_{y^{(h)}} = \int dF_{y^{(h)}}\big(k_1^{(h)},\ldots,k_{2+n_h}^{(h)}\big)
\,\Theta_{Y^{(h)}}\big(k_1^{(h)},\ldots,k_{2+n_h}^{(h)}\big)
\,\myScript{W}_{y^{(h)}}\big(k_1^{(h)},\ldots,k_{2+n_h}^{(h)}\big)
~.
\end{equation}
%%%%%%%%%%%%%%%%%%%%%%%%%%%%%%%%%%%%%%%%
%
These are estimated during the optimization of the distributions $F_{y^{(h)}}$, which happens as if they concerned single-parton scattering processes, with cuts $\Theta_{Y^{(h)}}$ that are chosen such that they cover enough phase space.
Each event in the event file gets a weight
%
%%%%%%%%%%%%%%%%%%%%%%%%%%%%%%%%%%%%%%%%
\begin{equation}
W = 
\frac{\myScript{S}_{\mathrm{MPI}}}{\sigma_{\mathrm{eff}}}
\prod_{h=1}^{\NMPI}\left[
\myScript{W}_{y^{(h)}}\big(k_1^{(h)},\ldots,k_{2+n_h}^{(h)}\big)
\,\frac{\sum_{y'\in Y^{(h)}}\sigma_{y'}}{\sigma_{y^{(h)}}}\right]
~.
\end{equation}
%%%%%%%%%%%%%%%%%%%%%%%%%%%%%%%%%%%%%%%%
%
The PDFs used to create the event file are of the single-parton type.
At the moment, it is not possible to use PDFs which depend in non-factorizable manner on the variables of more than one parton.
Such dependencies can be included after the event file has been created, by re-weighting the events.
The event file provides the value of $\myScript{P}_{y_1(h)}\myScript{P}_{y_2(h)}$ for each $Y^{(h)}$, and non-factorizable PDFs can be included by multiplying the event weight with
%
%%%%%%%%%%%%%%%%%%%%%%%%%%%%%%%%%%%%%%%%
\begin{equation}
\frac{\myScript{D}\Big(k_1^{(1)},k_2^{(1)};\ldots;k_1^{(\NMPI)},k_2^{(\NMPI)}\Big)}
{\prod_{h=1}^{\NMPI}\myScript{P}_{y_1(h)}\Big(k_1^{(h)}\Big)\myScript{P}_{y_2(h)}\Big(k_2^{(h)}\Big)}
~,
\end{equation}
%%%%%%%%%%%%%%%%%%%%%%%%%%%%%%%%%%%%%%%%
%
where $\myScript{D}$ is the desired (arbitrary) multi-parton PDF.

In order to studie for example both the contributions from single-parton scattering (SPS) and double-parton scattering (DPS) to a process, the user must generate separate event files for each.
The (differential) cross sections from each can simply be added together.
\KaTie\ does not provide the means to generate ``mixed'' event files.
\section{Usage\label{Usage}}
\KaTie\ uses many Fortran 2003 features, but not all of them (for example not parametrized derived types).
It requires a compiler which provides at least the Fortran 2003 features that gfortran-4.6 provides.
Furthermore, it requires Python-2.x with x$\geq$6 or Python-3.x, and Bash.
The program can be obtained from
\myvm{http://bitbucket.org/hameren/katie/downloads}
It comes as a \mytt{.zip} file, that can be extracted with \mytt{unzip}.
For the following, we refer to the directory where \KaTie\ is installed, that is the directory created by the unzipping procedure%
% and contains the file \mytt{settings.py}
, with \mytt{\$KaTie}.
Before \KaTie\ can be used, the user must download \avhlib, which can be obtained from
\myvm{http://bitbucket.org/hameren/avhlib/downloads}
and must also be unzipped.
We refer to the resulting directory with \mytt{\$AVHLIB}.
Next, the user must edit the file \mytt{\$KaTie/settings.py}, and set the path to the \avhlib-directory (so the value of \mytt{\$AVHLIB}), the path to the LHAPDF directory (the directory of the library file), and the Fortran compiler.
LHAPDF is also required if no collinear PDFs are used, in order to provide the running coupling constant.
It is advised to use at least version 6.x.
%
%Many platforms have a standardized possibility to install LHAPDF, for example it is included in the Arch User Repository (AUR) for Arch Linux.
%
If the user has {\sc TMDlib} available and wishes to use it, then they also need to set the path to the directory where its library file is, and the path to the directory where the library file of the GNU Scientific Library (GSL) is.

After finishing the settings file, the user must execute
\myvm{\$ \$KaTie/config.py lib}
in order to create a library.
This also configures \avhlib.
The easiest way to proceed is to choose an example in the directory \mytt{\$KaTie/examples} that is closest to the user's wishes, and use this as a starting point.
Let us say it is \mytt{pp\_to\_4j}.
The user must choose a non-existent directory, from now on referred to as \mytt{\$project}, and execute
\myvm{\$ \$KaTie/work.sh pp\_to\_4j \$project}
This will create \mytt{\$project} with some files and a symbolic link to the run-script inside.
In this example, there are two input files: one for single-parton scattering (SPS) and one for double-parton scattering (DPS).

If, for some reason, the library needs to be re-compiled at some point, this can be achieved with
\myvm{\$ \$KaTie/run.sh clean\\
      \$ \$KaTie/run.sh lib}
This is the run-script that is linked in \mytt{\$project}, so the re-compilation can also be invoked from there.

\subsection{Input file}
The input file consist of lines with the structure \mytt{keyword = value}.
The `\mytt{=}' must be separated from \mytt{keyword} and \mytt{value} by at least one space.
Words in a line in the input file must always be separated by at least one space, also when ``words'' consist of a single character, like `\mytt{=}'.
As a first example of a keyword, we mention the \mytt{include} statement, with which the lines of another input file can be included:
\myvm{include = someInputFile}
We will continue with DPS, because it requires slightly more explanation.

\subsubsection{Processes}
The first parameter in \mytt{\$projects/input\_dps}
\myvm{Ngroups = 2}
sets the number of scattering groups.
This is 1 for SPS, 2 for DPS etc.
In this case the number of final-state particles must be set for each of these, thus 
the 2 numbers on the right in the line
\myvm{Nfinst = 2 2}
Next, the contributing processes are set.
The user has to compose the list of processes themselves.
Although this may not seem most convenient, it does give the user full flexibility in which processes to include.
A process line has four keywords, with values that can consist of a variable number of items.
Each keyword and each item counts as a ``word'' that has to be separated from other ``words'' with at least one space.
An example of a process line is 
\myvm{process = q q\textasciitilde\ -> r r\textasciitilde\ \  factor = Nf-1\ \   groups = 1 2\ \   pNonQCD = 0 0 0}
It refers to the parton-level process $q\bar{q}\to q'\bar{q}'$.
Using the symbols \mytt{q} and \mytt{r} to refer to quarks implies summation over initial-state combinations of quarks that have the same matrix element within pure QCD.
This would be incorrect if electro-weak interactions are involved, and then the quarks should be denoted \mytt{u} for up-type quarks and \mytt{d} for down-type quarks.
Then there is still a summation over combinations that have equal matrix elements.
All combinations are explicitly given in \Appendix{quarksums}.
This is the default treatment.
If no summation is desired, then a parameter named \mytt{partlumi} must be set as follows:
\myvm{partlumi = individual}
Summation over final-state quarks with equal matrix elements is achieved via the \mytt{factor} parameter in the process line.
There, \mytt{Nf} is the number of flavors \mytt{Nflavors}, which is set later.
It is up to the user to put the correct factor.
Simple arithmetics is naturally interpreted, so it can be set to for example \mytt{Nf*(Nf-1)/2} for more complicated processes (here no spaces).
The next parameter, \mytt{groups}, indicates to which groups the process contributes. %
In the case of DPS for 4 jets, all $2\to2$ processes contribute to both groups, but for the production of $\mu^+\mu^-\,jj$ for example this would not be the case.
Then, the $2\to2$ processes with the muons in the final state would only contribute to one group, and the processes with partons in the final state to the other group.
%
%Then one would have for example the lines
%
%\myvm{process = q q\textasciitilde\ -> mu+ mu-\ \  factor = 1\ \   groups = 1\ \   pNonQCD = 2 0 0\\
%      process = q q\textasciitilde\ -> r r\textasciitilde\ \  factor = Nf-1\ \   groups = 2\ \   pNonQCD = 0 0 0}
%
The last parameter in the process line, \mytt{pNonQCD}, indicates the power of the electro-weak coupling, the Higgs-gluon coupling, and the Higgs-photon coupling respectively.
This is relevant again for example for $pp\to\mu^+\mu^-\,jj$, to which both $\myScript{O}\big(\alpha_\mathrm{S}^2\alpha_\mathrm{EW}^2\big)$ and $\myScript{O}\big(\alpha_\mathrm{EW}^4\big)$ Feynman graphs contribute, while one would like to exclude the latter.
The names of all possible particles are
\myvm{ve\ \ \ ve\textasciitilde\ \ \ e-\ \ \ e+\ \ \ u\ u\textasciitilde\ d\ d\textasciitilde\\
      vmu\ \  vmu\textasciitilde\ \  mu-\ \  mu+\ \  c\ c\textasciitilde\ s\ s\textasciitilde\\
      vtau    vtau\textasciitilde\   tau-\   tau+\   t\ t\textasciitilde\ b\ b\textasciitilde\\
      g\ H\ A\ Z\ W+\ W-}
They should all be obvious, except maybe the neutrinos which start with \mytt{v}, and the photon \mytt{A}.

\subsubsection{\label{subsubPDFs}PDFs}
The next parameter sets the pdf set from LHAPDF, for example
\myvm{lhaSet = MSTW2008nlo68cl}
It needs to be set also for off-shell initial-state partons, because \KaTie\ takes $\alpha_\mathrm{S}$ from there.
The line
\myvm{offshell = 0 0}
determines that no partons are off-shell.
The other possibilities are \mytt{0 1}, \mytt{1 0}, and \mytt{1 1}.

In order to use a TMD PDF set from {\sc TMDlib}, the associated keyword from Table~1 in~\cite{Hautmann:2014kza}, for example \mytt{KS-2013-linear}, can be set with
\myvm{TMDlibSet = KS-2013-linear}
The initial state(s) marked as off-shell will then be treated with the PDFs from that set.

Alternatively, the user can provide TMD PDFs as grids directly.
They must be given in the form of files consisting of three or four columns, containing
$$
\ln(x) \quad  \ln\big(|\mathbf{k}_T|^2\big) \quad xf(x,|\mathbf{k}_T|)
$$
or
$$
\ln(x) \quad  \ln\big(|\mathbf{k}_T|^2\big) \quad \ln\big(\mu^2\big) \quad xf(x,|\mathbf{k}_T|,\mu)
$$
if the pdf also depends on the factorization scale.
The scale dependence, \ie\ the fact that the file has four collumns, does not have to be indicated anywhere, and is recognized automatically.
\KaTie\ interpolates the grids itself.
The directory where \KaTie\ can find the grids is indicated in the input file by
\myvm{tmdTableDir = /home/user/projects/tmdgrids/}
where here of course an example path is given.
The actual grid file must be indicated for each parton separately with
\myvm{%
tmdpdf = g  gluon.dat\\\noindent
tmdpdf = u  uQuark.dat\\\noindent
tmdpdf = u\textasciitilde\  uBar.dat
}
etc., where the file names are examples again.
The path can be changed (or rather updated) between lines indicating the files, in case the files are distributed over several directories.
In the case of $k_T$-factorization, a TMD for the gluon {\em must always\/} be provided, also if the user happens to only want to study quark-initiated processes.
Setting the keyword \mytt{TMDlibSet} overrules the grids.
The number of active flavors is set with
\myvm{Nflavors = 5}
and in case of DPS, the value of $\sigma_{\mathrm{eff}}$ is set with
\myvm{sigma\_eff = 15d6}
in units of nanobarn.

\subsubsection{Kinematics and cuts\label{kinematicsAndCuts}}
The center-of-mass energy of the scattering is set in GeV with
\myvm{Ecm = 7000}
The phase space pre-sampler needs information about the typical value of the softest scale, like for example a minimum $p_T$, which is set in GeV with
\myvm{Esoft = 20}
This number {\em is not a cut-off\/} and only influences the efficiency of the optimization, as explained in \Appendix{Esoft}.
It must be larger than zero.
The actual phase space cuts are set explicitly in the input file.
\myvm{cut = \{deltaR|2,4|\} > 0.5}
sets the minimum value of $\Delta R$ between final-state particle $2$ and $4$.
At tree-level, to which \KaTie\ currently is restricted, this corresponds to the value of $R$ in a jet algorithm.
The particle numbers above refer to the order as given in the process list.
\myvm{cut = \{pT|i|j,k,...\} > 50}
sets the minimum $p_T$ in GeV for the $i$-th final-state in the $p_T$-ordered list of final-state number $j,k,...$.
For example, if the process is, for some unpractical reason, given as $pp\to j\,\mu^+\,j\,\mu^-\ j\,j$, then the jets are coming from final-state items \mytt{\{1,3,5,6\}}.
One of them will have the second-highest jet $p_T$, and demanding that it is at least $50$GeV is done by
\myvm{cut = \{pT|2|1,3,5,6\} > 50}
A minimum $p_T$ for final-state $3$ directly can be set with \mytt{\{pT|3|\}}, and a maximum $p_T$ can be set with \mytt{<}.
\myvm{cut = \{mass|1+3+4|\} > 100}
sets a minimum for the invariant mass of the sum of final-state momenta $1$, $3$, and $4$.
Similarly, other variables also can take arguments that consist of sums, \eg\ the $p_T$ of the sum of final-state momenta $2$ and $3$ is \mytt{\{pT|2+3|\}}.
Other possible variables are \mytt{rapidity}, \mytt{pseudoRap, \mytt{deltaPhi}}, and \mytt{ET} (also called the transverse mass):
$E_T = \sqrt{p_T^2+p^2} = \sqrt{(E-p_z)(E+p_z)}$~.
%
%%%%%%%%%%%%%%%%%%%%%%%%%%%%%%%%%%%%%%%%
%\begin{equation}
%E_T = \sqrt{p_T^2+p^2} = \sqrt{(E-p_z)(E+p_z)}
%~.
%\end{equation}
%%%%%%%%%%%%%%%%%%%%%%%%%%%%%%%%%%%%%%%%
%
The meaning of the other variables should be obvious.
The values of \mytt{rapidity} and \mytt{pseudoRap} can be negative and the variables do not refer to the absolute value.
On the other hand, \mytt{deltaPhi} is positive between $0$ and $\pi$.

It is assumed that the factorization scale and the renormalization scale are identical in \KaTie.
It goes both into the PDFs and $\alpha_{\mathrm{S}}$, and is set with for example
\myvm{scale = (\{pT|1|\}+\{pT|2|\}+\{pT|3|\}+\{pT|4|\})/2}
No spaces are allowed in the expression on the right-hand side.
All variables mentioned above are allowed.
Numerical constants and the arithmetic operations \mytt{+}, \mytt{-}, \mytt{*}, \mytt{/}, \mytt{**}, as well as parentheses are allowed. 
In the case of DPS, one would like to have separate scales for the different groups.
This can be achieved with for example
\myvm{%
scale = entry 1 (\{pT|1|\}+\{pT|2|\})/2\\
scale = entry 2 (\{pT|3|\}+\{pT|4|\})/2%
}
In the case of DPS, cuts and scale also have to be given for each group separately for the phase space optimization.
The phase spaces are optimized independently, as if they were SPS processes.
To set these, the lines in the input file need, to look like
\myvm{%
cut = group 1 \{pT|2|1,2\} > 50 \\\noindent
scale = group 1  (\{pT|1|\}+\{pT|2|\})/2 \\\noindent
cut = group 2 \{pT|2|1,2\} > 50 \\\noindent
scale = group 2  (\{pT|1|\}+\{pT|2|\})/2
}
etc..
Notice that the particle numbering for \mytt{group 2} is also $1,2$ now, as if it was an SPS process.

As mentioned earlier, the denominator of the flux factor in \Equation{fluxfactor} is not prescribed by a specific factorization theorem.
It is argued in \cite{Nefedov:2015ara} that within the framework of Kimber, Martin, and Ryskin~\cite{Kimber:1999xc} the appropriate denominator is the generalization of the one in collinear factorization, so
%
%%%%%%%%%%%%%%%%%%%%%%%%%%%%%%%%%%%%%%%%
\begin{equation}
\flux(k_1,k_2)
=
\frac{\theta\big((k_1+k_2)^2\big)}{8k_1^0k_2^0}
~.
\end{equation}
%%%%%%%%%%%%%%%%%%%%%%%%%%%%%%%%%%%%%%%%
%
This is the default in \KaTie.
The one of \Equation{fluxfactor} can be set with the option
\myvm{flux factor = textbook}

\subsubsection{Model parameters}
Particle masses and widths are set in GeV with
\myvm{%
mass = Z   91.1882  2.4952  \\\noindent
mass = W   80.419   2.21    \\\noindent
mass = H  125.0     0.00429 \\\noindent
mass = t  173.5
}
The absence of a value for the width, like for the top quark above, implies it is set to zero.
Other particles are massless by default, but can be given a mass and width too.
The user can indicate which interactions are active with the lines
\myvm{%
switch = withQCD   Yes \\\noindent
switch = withQED   No  \\\noindent
switch = withWeak  No  \\\noindent
switch = withHiggs No  \\\noindent
switch = withHG    No  \\\noindent
switch = withHA    No
}
where the last two refer to the effective Higgs-gluon and Higgs-photon interactions (explicit expressions for the vertices can for example be found in~\cite{vanDeurzen:2013rv}).
These and \mytt{withHiggs} are switched off by default, while the others are switched on.
The electro-weak coupling is fixed, and can be set with for example
\myvm{coupling = alphaEW 0.00794}
Alternatively, the value of Fermi's constant $G_\mathrm{F}$ can be set with for example
\myvm{coupling = Gfermi 1.16639d-5}
The electo-weak coupling is then set to
$\alpha_\mathrm{EW}=G_\mathrm{F}\frac{\sqrt{2}}{\pi}m^2_W\big(1-\frac{m^2_W}{m^2_Z}\big)$.
The value of the effective Higgs-gluon coupling is set with
\myvm{coupling = Higgs-gluon -0.431d-3}
This is the amplitude-level effective coupling and should contain everything except the overall factor of $\alpha_{\mathrm{S}}$, so the value of $g_{\mathrm{rest}}$ in $g_h = \alpha_{\mathrm{S}}\,g_{\mathrm{rest}}$.
Possible higher powers of $\alpha_{\mathrm{S}}$ can only be included as fixed numbers in $g_{\mathrm{rest}}$ and cannot run.
For \mytt{Higgs-photon} the user provides the whole, fixed, value of the amplitude-level effective coupling.

\subsubsection{Optimization parameters}
If there are many final-state particles, the sum over helicities becomes unnecessarily time consuming, and the user should choose to sample over helicities instead:
\myvm{helicity = sampling}
Other possible values are \mytt{sum} and \mytt{polarized}.
The latter chooses the continuous sampling method of \cite{Draggiotis:1998gr}.
The number of events to be spent on the optimization of the pre-sampler also needs to be set in the input file.
\myvm{Noptim = 100,000}
sets the number of non-vanishing-weight events to a hundred thousand.
This will be discussed in more detail in the following.

\subsection{Optimization stage\label{optimizationstage}}
Event generation happens in two stages.
During the first stage, the pre-sampler is optimized for each process given in the input file separately.
Executing
\myvm{\$ \$project/run.sh prepare \$project/input\_sps \$project/trial01}
will create a directory \mytt{\$project/trial01} and inside a directory will be created for each process given in the input file \mytt{input\_sps}.
The name \mytt{trial01} is just an example.
Executing
\myvm{\$ \$project/trial01/optimize.sh}
will start the optimization of all processes.
If there are very many processes, one might want the optimization to happen in batches, and by executing for example
\myvm{\$ \$project/trial01/optimize.sh Nparallel=4}
only $4$ optimizations will run simultaneously until all have been performed.
One can also select processes to be optimized, \eg
\myvm{\$ \$project/trial01/optimize.sh Nparallel=4 proc=1,7,13,24,25}
The progress can be monitored by viewing the output files in each process directory, for example with
\myvm{\$ tail -f \$project/trial01/proc*/output}
The final precision should not be more than a few percent.
For example
\myvm{%
MESSAGE from Kaleu stats:\ Ntot = 38,887\\
MESSAGE from Kaleu stats:\ + \ 25,600 (.13072883+/-.00132728)E+06  1.015\% 
}
\noindent
means that 38887 events were generated, of which 25600 passed the cuts, leading to an estimated cross section of $0.1307\times10^6\mathrm{nb}$ with an estimated statistical error of 1.015\%. 
The \mytt{+} in front of \mytt{25,600} means that it concerns positive-weight events.
In case the optimization of a process does not seem to converge, the user can try to increase the number of events for that process and/or change the random number seed.
Say the user wants to rerun  process number 3  with  seed=273846  and with Noptim=200000.
Then the user needs to execute
\myvm{%
\$ prefix=\$project/trial01/proc03 \\\noindent
\$ \$prefix/main.out seed=273846 Noptim=200000 > \$prefix/output \&
}
Alternatively, the user can edit the appropriate lines in \mytt{\$project/trial/optimize.sh}.
For any seed, the result of the optimization will be an unbiased, but not necessarily efficient, phase space pre-sampler.
The seed can be different for every process.
The only rule is that once you start to generate event files, the pre-sampler must be fixed, and you MUST NOT rerun the optimization.  

\subsection{Event generation}
The second stage is the actual event generation, and happens after the optimization is finished.
The user may just execute
\myvm{%
\$ prefix=\$project/trial01 \\\noindent
\$ nohup \$prefix/main.out seed=732415 > \$prefix/output1 \& \\\noindent
\$ nohup \$prefix/main.out seed=232984 > \$prefix/output2 \&
}
etc.\ for several different random number seeds.
So-called ``raw'' event files will be produced in \mytt{\$project/trial01} with the names \mytt{raw732415.dat},  \mytt{raw232984.dat} etc.
They contain weighted events that form a partially un-weighted collection of events from all that are generated by the pre-sampler.
The number of events and the rate of fluctuation of their weights can be steered to some degree, as explained in \Section{formalism}.
By default, of the order of $10^5$ events are accepted when the cross section is estimated to a statistical precision of $0.001=0.1\%$.
These numbers can be changed by providing optional arguments.
For example
\myvm{%
\$ nohup \$prefix/main.out Nevents=1e6 precision=0.01 seed=732415 \textbackslash \\\noindent> \$prefix/output1 \&
}
will try to accept of the order of $10^6$ events until a statistical precision of $0.01=1\%$ is reached.
In this case, of course, the events will be of ``lower quality'' in the sense that their weights will fluctuate more.

The standard output of a Monte Carlo run, sent to output files in the examples above, will, after some initialization, consist of lines looking as follows:
\myvm{\ \ \ 1,579,644\ \ \ \ \ \ 19,600\ \ (.28868124+/-.01110617)E+03\ \ \ 3.847\%}
These data are the number of generated events so far, the number of events that passed the cuts, the estimate of the cross section in nb with error estimate, and the estimated relative error.

In order to create a LHEF \mytt{\$project/trial01/eventfile.dat}, the user must execute
\myvm{\$ \$project/run.sh lhef \$project/trial01/raw*}
This will use all available raw files.
The user may also select some by listing them separately in the above command.
This command may be executed before the event generations have finished.
The event file will be created in the directory where \mytt{run.sh} is executed.
Alternatively, in order to create an event file in a custom format that can be processed with \KaTie\ further, the user may execute
\myvm{\$ \$project/run.sh merge \$project/trial01/raw*}
An ASCII file \mytt{\$project/trial01/eventfile.dat} will be created.
After obvious information in the header, the events are listed.
Each event block starts with, for example
\myvm{EVENT WEIGHT:  0.5778970106138136E+09}
where the floating point number is the weight value $W$ assigned to the event.
The weights are normalized such that
%
%%%%%%%%%%%%%%%%%%%%%%%%%%%%%%%%%%%%%%%%
\begin{equation}
\frac{\sum_\mathrm{events}W}{\sum_\mathrm{events}1} = \textrm{total cross-section}
~.
\nonumber
\end{equation}
%%%%%%%%%%%%%%%%%%%%%%%%%%%%%%%%%%%%%%%%
%
The next line contains a single integer, referring to the process corresponding to the event.
The process numbering is given in the header of the event file.
The next lines consist of 5 floating point numbers and 2 integers, containing the momenta and the color flow of the event:
%
%%%%%%%%%%%%%%%%%%%%%%%%%%%%%%%%%%%%%%%%
\begin{equation}
E \qquad p_x \qquad p_y \qquad p_z \qquad E^2-p_x^2-p_y^2-p_z^2
\qquad\textrm{color} \qquad \textrm{anti-color}
\nonumber
\end{equation}
%%%%%%%%%%%%%%%%%%%%%%%%%%%%%%%%%%%%%%%%
%
Initial-state momenta have a negative value of the energy $E$.
Then follows a line consisting of 4 floating point numbers, containing the value of
\begin{center}
matrix element \qquad parton luminosity \qquad $\alpha_{\mathrm{S}}$ \qquad scale
\end{center}
The matrix element is not averaged regarding the initial-state partons.
The parton luminosity is just the product of the PDFs, as $x_1f_1\,x_2f_2$, not $f_1f_2$.
This number is also included in the LHEF, in lines starting with \mytt{\#pdf1pdf2}~.
For multi-parton scattering, event blocks are repeated for a single event weight.
For MPI, event files can only be created in the custom format for now, and not in the LHEF format.

\subsection{Creation of histograms}
The custom format event file can be used to make histograms with \KaTie.
The user may use the Fortran program \mytt{read\_event\_file.f90} in the directory \mytt{\$project} as a starting point.
It can be compiled and executed with
\myvm{%
\$ \$project/run.sh compile \$project/read\_event\_file.f90 \\\noindent
\$ \$project/read\_event\_file.out \$project/trial01/eventfile.dat
}
The program uses the Fortran module \mytt{read\_events\_mod} which provides several variables, functions and subroutines, as well as two types to create one-dimensional and two-dimensional histograms.
They are listed at the beginning of \mytt{read\_event\_file.f90}.
The program has a declaration block, an initialization block to set the bins for the histograms etc., a block that runs through the event file to collect the data, and a block to write the histograms to files with user-chosen file names.
An array of histograms can, for example, be declared as
\myvm{type(histo\_1d\_type) :: h\_pT(1:4)}
and the entries can be initialized with
\myvm{do ii=1,4\\
   call h\_pT(ii)\%init( left=0d0 ,right=200d0 ,Nbins=100 )\\
enddo
}
It is also possible to set bins explicitly, \eg
\myvm{call h\_pT(1)\%init([0d0,50d0 ,50d0,100d0 ,100d0,150d0 ,150d0,200d0])}
will set four bins of size 50GeV (the entries in the array may also be single precision).
An example of the available variables is the array \mytt{pFnst(0:3,:)}, containing the final-state four-momenta of the event (\mytt{pFnst(0,i)} is the energy of final-state momentum \mytt{i}), and an example of the available functions is \mytt{pTrans}, returning the size of the transverse momentum of a four-momentum.
Then,
\myvm{%
do ii=1,4\\
  pT(ii) = pTrans(pFnst(0:3,ii))\\
enddo}
will fill the array \mytt{pT} with the value of the size of the transverse momenta of the first four final-state momenta
These can be ordered with
\myvm{call sort\_big2small( pTordered ,pT ,Njet )}
which alters \mytt{pT}, but also returns the integer array \mytt{pTordered} containing the associated permutation: \mytt{pT(i)=pTrans(pFinst(0:3),pTordered(i))}.
So \mytt{pT(1)} will be the largest, \mytt{pT(2)} the next to hardest etc.
The histograms are filled with
\myvm{%
do ii=1,4\\
  call h\_pT(ii)\%collect( pT(ii) ,eventWeight )\\
enddo
}
where \mytt{eventWeight} is the event weight.
Finally, the histograms are written to files with
\myvm{%
do ii=1,4\\
  call h\_pT(ii)\%write('pT'//numeral(ii)//'.hst')\\
enddo
}
The character array \mytt{numeral(0:9)} contains the numbers $0$ to $9$ as single characters, so the filenames will be \mytt{pT1.hst}, \mytt{pT2.hst}, etc.
The files are written to the directory where the user executed \mytt{read\_event\_file.out}.
They consist of four columns, containing the left bin-border, right bin-border, the value, and the statistical error estimate.
Histograms can be plotted with, for example, gnuplot via
\myvm{plot 'pT1.hst' using (\$1+\$2)/2:3 with boxes}

\section{\label{Sec:summary}Summary}
\KaTie, a program for parton-level generation of events with initial-states that can have non-vanishing transverse momentum components, was presented.
It provides all necessary ingredients, including off-shell matrix elements and an efficient importance sampler, except the transverse momentum dependent parton density functions.
The latter can be provided in the form of hyper-rectangular grids which \KaTie\ will automatically interpolate, or by {\sc TMDlib}.
Events are produced in the Les Houches Event File format, or in a custom format with which distributions of arbitrary observables can be produced with tools also provided by \KaTie.
Finally, a convenient environment is available to perform calculations involving multi-parton interactions.

\subsection*{Acknowledgments}
The author would like to thank Marcin Bury, Hannes Jung, Mirko Serino, and Krzysztof Bo{\.z}ek for their help in beta testing the program.
Also, the author would like to thank Maxim Nefedov and Vladimir Saleev for valuable feedback.
This work was supported by grant of National Science Center, Poland, No.\ 2015/17/B/ST2/01838.

%\bibliography{bibliography}{}\bibliographystyle{JHEP}
%\bibliography{refs}{}\bibliographystyle{JHEP}
%\input{8_references.tex}

\begin{appendix}
\section{PDF sums\label{quarksums}}
In the following, the indices $1,2$ refer to initial-state $1$ and $2$.
Each left-hand-side below implies the sum on the right-hand side, because for each term on the right-hand side the numerical value of the matrix element is identical.
We consider the $5$-flavor case.
For the $4$-flavor case $b$ and $\bbar$ quarks are removed, for the $3$-flavor case also $c$ and $\cbar$ quarks are removed, and for the $2$-flavor case also $s$ and $\sbar$ quarks are removed.
In case of no electro-weak interactions, there are the following initial-state cases, which all need to be included for example for the processes $pp\to\mathrm{jets}$:
%
%%%%%%%%%%%%%%%%%%%%%%%%%%%%%%%%%%%%%%%%
\begin{align}
 \lhs{g}{g}    &: g_1 g_2 \\
 \lhs{g}{q}    &: g_1(u_2+\ubar_2+c_2+\cbar_2+d_2+\dbar_2+s_2+\sbar_2+b_2+\bbar_2) \\
 \lhs{q}{g}    &: (u_1+\ubar_1+c_1+\cbar_1+d_1+\dbar_1+s_1+\sbar_1+b_1+\bbar_1)g_2 \\
 \lhs{q}{q}    &: u_1 u_2+\ubar_1\ubar_2 +c_1 c_2+\cbar_1\cbar_2
                 +d_1 d_2+\dbar_1\dbar_2 +s_1 s_2+\sbar_1\sbar_2 +b_1 b_2+\bbar_1\bbar_2 \\
 \lhs{q}{\qBAR}&: u_1\ubar_2+\ubar_1 u_2 +c_1\cbar_2+\cbar_1 c_2
                 +d_1\dbar_2+\dbar_1 d_2 +s_1\sbar_2+\sbar_1 s_2 +b_1\bbar_2+\bbar_1 b_2
%\\
\end{align}
\begin{align}
  \lhs{q}{r}&:(u_1+\ubar_1)(\phantom{u_2+\ubar_2+}c_2+\cbar_2+d_2+\dbar_2+s_2+\sbar_2+b_2+\bbar_2)\notag\\
    &+(c_1+\cbar_1)(u_2+\ubar_2+\phantom{c_2+\cbar_2+}d_2+\dbar_2+s_2+\sbar_2+b_2+\bbar_2)\notag\\
    &+(d_1+\dbar_1)(u_2+\ubar_2+c_2+\cbar_2+\phantom{d_2+\dbar_2+}s_2+\sbar_2+b_2+\bbar_2)\\
    &+(s_1+\sbar_1)(u_2+\ubar_2+c_2+\cbar_2+d_2+\dbar_2+\phantom{s_2+\sbar_2+}b_2+\bbar_2)\notag\\
    &+(b_1+\bbar_1)(u_2+\ubar_2+c_2+\cbar_2+d_2+\dbar_2+s_2+\sbar_2\phantom{+b_2+\bbar_2})\notag
\end{align}
In the case electro-weak interactions are involved, the foregoing is incorrect because the matrix elements are not equal for all those combination.
Then, the correct decomposition is as follows:
\begin{align}
 \lhs{g}{u}    &: g_1(u_2+c_2)   &      
 \lhs{g}{d}    &: g_1(d_2+s_2+b_2) \\
 \lhs{g}{\uBAR}&: g_1(\ubar_2+\cbar_2) &
 \lhs{g}{\dBAR}&: g_1(\dbar_2+\sbar_2+\bbar_2) \\
 \lhs{u}{g}    &: (u_1+c_1)g_2   &      
 \lhs{d}{g}    &: (d_1+s_1+b_1)g_2 \\
 \lhs{\uBAR}{g}&: (\ubar_1+\cbar_1)g_2 &
 \lhs{\dBAR}{g}&: (\dbar_1+\sbar_1+\bbar_1)g_2
\end{align}
\begin{align}
 \lhs{u}{u}        &: u_1 u_2 +c_1 c_2&
 \lhs{d}{d}        &: d_1 d_2 +s_1 s_2 +b_1 b_2 \\
 \lhs{\uBAR}{\uBAR}&: \ubar_1\ubar_2 +\cbar_1\cbar_2 &
 \lhs{\dBAR}{\dBAR}&: \dbar_1\dbar_2 +\sbar_1\sbar_2 +\bbar_1\bbar_2 \\
 \lhs{u}{\uBAR}    &: u_1 \ubar_2 +c_1 \cbar_2 &
 \lhs{d}{\dBAR}    &: d_1 \dbar_2 +s_1 \sbar_2 +b_1 \bbar_2 \\
 \lhs{\uBAR}{u}    &: \ubar_1u_2 +\cbar_1c_2 &
 \lhs{\dBAR}{d}    &: \dbar_1d_2 +\sbar_1s_2 +\bbar_1b_2 
\end{align}
\begin{align}
 \lhs{u}{d}        &: (u_1+c_1)(d_2+s_2+b_2) &
 \lhs{u}{\dBAR}    &: (u_1+c_1)(\dbar_2+\sbar_2+\bbar_2) \\
 \lhs{\uBAR}{d}    &: (\ubar_1+\cbar_1)(d_2+s_2+b_2) &
 \lhs{\uBAR}{\dBAR}&: (\ubar_1+\cbar_1)(\dbar_2+\sbar_2+\bbar_2) \\
 \lhs{d}{u}        &: (d_1+s_1+b_1)(u_2+c_2) &
 \lhs{\dBAR}{u}    &: (\dbar_1+\sbar_1+\bbar_1)(u_2+c_2) \\
 \lhs{d}{\uBAR}    &: (d_1+s_1+b_1)(\ubar_2+\cbar_2) &
 \lhs{\dBAR}{\uBAR}&: (\dbar_1+\sbar_1+\bbar_1)(\ubar_2+\cbar_2)
\end{align}
%%%%%%%%%%%%%%%%%%%%%%%%%%%%%%%%%%%%%%%%

\section{The meaning of \texttt{Esoft}\label{Esoft}}
Pre-sampling is performed with {\sc Kaleu}~\cite{vanHameren:2010gg}, which constructs phase space points from invariants and angles that are generated following pre-defined probability distributions augmented with adaptive grids, following the method of~\cite{vanHameren:2007pt}\footnote{\avhlib\ employs {\sc RANLUX}~\cite{James:1993np} for the generation of pseudo random numbers.}.
The generation of the invariants is such that it mimics the behavior of the squared matrix element as function of the invariants.
For example, if the final-state momenta $p_i,p_j$ belong to on-shell gluons, then the squared matrix element behaves as $1/s_{ij}$ where $s_{ij}=(p_i+p_j)^2$.
The singularity at $s_{ij}=0$ is protected by the phase space cuts, but the squared matrix element will still show a steep behavior as function of $s_{ij}$.
Therefore, the predefined density for the invariant is
%
%%%%%%%%%%%%%%%%%%%%%%%%%%%%%%%%%%%%%%%%
\begin{equation}
P_{ij}(s_{ij})
=
\frac{\theta\big(\Ecm^2-s_{ij}\big)}{1+\log\big(\Ecm^2/\Esoft^2\big)}\left[
\frac{\theta\big(\Esoft^2-s_{ij}\big)}{\Esoft^2}
+
\frac{\theta\big(s_{ij}-\Esoft^2\big)}{s_{ij}}
\right]
~.
\end{equation}
%%%%%%%%%%%%%%%%%%%%%%%%%%%%%%%%%%%%%%%%
%
It increases for decreasing $s_{ij}$ until $s_{ij}=\Esoft^2$, from where it stays constant, ensuring the coverage of the whole phase space.
The exact value of $\Esoft$ does not matter too much, since also a bad choice will be corrected by the adaptive grid.
A good choice will, however, help in the efficiency.
\end{appendix}

\end{document}